\documentclass{aa}
\usepackage{txfonts}
\usepackage{graphicx}
\usepackage{natbib}
\bibpunct{(}{)}{;}{a}{}{,}        

\newcommand{\fo}{\ensuremath{f^\mathrm{o}}}
\newcommand{\fe}{\ensuremath{f^\mathrm{e}}}
\newcommand{\bcont}{\ensuremath{b_\mathrm{cont}}}
\newcommand{\rcont}{\ensuremath{r_\mathrm{cont}}}
\newcommand{\pq}{\ensuremath{P_Q}}
\newcommand{\pu}{\ensuremath{P_U}}
\newcommand{\pl}{\ensuremath{P_\mathrm{L}}}

\begin{document}

\title{Photometry and polarimetry of the nucleus of
comet 2P/Encke
\thanks{Based on observations performed under program 078.C-0509 
at the European Southern Observatory, Cerro Paranal, Chile}}
\author{H. Boehnhardt\inst{1} \and G.P. Tozzi\inst{2} \and S. Bagnulo\inst{3}
  \and K. Muinonen\inst{4} \and A. Nathues\inst{1} \and L. Kolokolova\inst{5}}

\institute{Max-Planck-Institute for Solar System Research, Max-Planck-Str. 2, 
D-37191 Katlenburg-Lindau, Germany (boehnhardt@mps.mpg.de)
\and
INAF Arcetri Observatory, Florence, Italy  
\and
Armagh Observatory, College Hill, Armagh BT61 9DG, Northern Ireland, U.K.
\and
University Observatory, Helsinki, Finland
\and
University of Maryland, College Park, Maryland, USA
 }

\date{Received today / Accepted tomorrow}

\abstract
{
Short-period comet 2P/Encke is a favorable target for
  studies of light scattering by bare cometary nuclei. These studies
  enable assessment of the nucleus size, 
  the material albedo, and the light-scattering properties of the cometary
  surface. 
}
{
We characterize the activity of the cometary
  coma and the physical properties of the nucleus of 2P/Encke such as its size,
  albedo, colors, spectral slope, surface roughness, porosity, and
  single-particle properties. 
}
{
Broadband imaging photometry and broadband and narrowband linear polarimetry
  is measured for the nucleus of 2P/Encke over the phase-angle range
  4 -- 28 deg. 
An analysis of the point spread function of the comet reveals only weak coma activity,
corresponding to a dust production of the order of 0.05 kg/s.
The nucleus displays a color
  independent photometric phase function of almost linear slope ($\beta$ = 
  0.050$\pm$0.004 mag/deg). The absolute $R$ filter magnitude at zero phase angle 
  is 15.05$\pm$0.05, and corresponds to an
  equivalent radius for the nucleus of 2.43$\pm$0.06km (adopted albedo of 
  0.047). The nucleus color $V - R$ is
  $0.47 \pm 0.07$, suggesting a spectral slope $S'$ of $11\pm 8$\,\%/100nm. 
  The phase function of linear polarimetry in the 
  $V$ and $R$ filter shows a widely color independent linear increase with 
  phase angle ($0.12\pm0.02$\,\%/deg).  We find discrepancies in the
  photometric and polarimetric parameters between 2P/Encke and other minor
  bodies in the solar system, which may indicate significant differences in the 
  surface material properties and light-scattering behavior of the bodies.
}
{
The linear polarimetric phase function of 2P/Encke presented here is the
  first ever measured for a cometary nucleus, and its analysis encourages
  future studies of cometary nuclei in order to characterize the 
  light-scattering behavior of comets 
  on firm empirical grounds and provide suitable input to a comprehensive 
  modeling of the light scattering by cometary surfaces.
}


\authorrunning{H. Boehnhardt et al.}
\titlerunning{Photometry and polarimetry of 2P/Encke}

\maketitle

\section{Introduction}\label{introduction}
After its discovery in 1786, the short-period comet 2P/Encke was
observed during almost 60 perihelion passages and is 
one of the most significantly (more than 40 papers in refereed journals) and 
longest (over 200 years) studied comets.  
With an orbital revolution period of about 3.3 years, it is closer to the Sun 
than typical Jupiter family comets (JFC). Levison et al. (2006) 
suggested that 2P/Encke may be a rare example of a comet in an orbit 
detached from Jupiter's immediate control (like JFCs are) 
due to gravitational interaction with the terrestrial planets. They also 
argued that
the nucleus may have been in a quiescent state for a long part of the
transition period into its current orbit.

A series of scientific papers dealt with the nucleus properties of
2P/Encke. Fernandez et al. (2000) measured an effective nuclear radius of
2.4$\pm$0.3km, an axial ratio of at least 2.6, a geometric albedo of 
$0.047\pm0.023$, which is in the typical range for comets, and a rather steep
photometric phase function with a linear slope of 0.06 mag/deg which was 
later confirmed by
Jewitt (2002). A wider range of values for the
effective radius (2.4--3.7 km) was determined from radar
echoes of the comet (Harmon and Nolan 2005). Colors of {\it V-R} = 0.39$\pm$0.06
and B-V = 0.73$\pm$0.06 (Lowry and Weissman 2007) indicated only moderate
spectral slope S' of the nuclear surface as found by Jewitt (2002) to be 
8.9$\pm$1.6 \%/100nm. The rotation
period was a matter of debate in various papers without a final conclusion (see
Belton et al. 2005 and references therein). Attempts to determine the
orientation of the rotation axis were made by Sekanina (1991) and
Festou and Barale (2000). Jewitt (2004) measured a slightly positive 
linear polarization of the nucleus at a single phase angle (polarization of
1.0--1.9 $\pm0.5$ \% at 22\degr) in
the visible. The activity profile of the nucleus appears to be unusual with a coma
being visible around aphelion (Fernandez et al. 2005), but no activity during its
inbound arc as close as 1.4 AU solar distance (Jewitt 2004).

The likely absence of activity combined with the predicted nucleus brightness 
enable us to attempt 
successfully (i.e. with sufficient signal-to-noise and without significant coma
contamination) polarization measurements of 2P/Encke. We report below on the
results of the phase-angle-resolved linear polarimetry and 
quasi-simultaneous broadband photometry  of this cometary nucleus in 
the visible wavelength range. The main aim of these measurements is to 
constrain the light-scattering properties of the nuclear surface, i.e. the 
single-scattering albedo and mean free path as a proxy 
for the typical grain
size to be obtained through modeling. Furthermore, we could test the
albedo-polarization relationship known from asteroids (Zellner and Gradie 1976,
Zellner et al. 1977, Lupishko and Mohamed 1996, Cellino et al. 1999), for a cometary
nucleus. Our results are the first ever linear 
polarimetric observations of a cometary nucleus over a wider phase-angle
range.


\section{Observations and Data Reduction}\label{observations}
The observations of comet 2P/Encke were performed at Unit Telescope 2 of the Very
Large Telescope observatory (VLT) at Cerro Paranal in Chile. The FORS1
instrument\footnote{see http://www.eso.org/instruments/ for technical details} 
was used in imaging and polarimetric modes with broadband Bessell $V$ and 
R and two narrowband filters of central wavelengths and widths 
485nm/37nm and 834nm/48nm, respectively; the filters are 
abbreviated below as ''\bcont'' for the blue and ''\rcont'' for the red one. 
The narrowband filters were selected from the
available set of FORS filters such that their transmission ranges contained
little to no contamination from emission bands of possible coma gases,
i.e. the narrowband filters cover mostly surface and dust reflected sunlight.
The \bcont\ and \rcont\  filters were only used during the polarimetric measurements at
the beginning of the program, since, due to the absence of significant coma 
contamination, we decided to maximize the signal-to-noise level by
exclusive usage of the broadband $V$ and $R$ filters from 28 Oct. 2006 onwards
(see Table \ref{obs_log}).

\subsection{Observations}\label{obsdetails} 
Service mode observations were the most appropriate choice given the
specific requirements of this program: The short duration (1-1.5h) of individual
observing runs; a frequent, but irregular run schedule to cover 
the orbit of 2P/Encke at about regular phase-angle differences between 4 and
30 deg phase angle; and acceptable sky conditions for the observations. 
To the advantage of a dense phase-angle coverage, we relaxed the requirements
for seeing (2'') and sky transparency (thin cirrus or better; in most cases the measurements
were performed with clear sky) for our program. The comet was positioned at
the center of the instrument field of view, a location that placed it
automatically in the center of the central strip of the FORS1 polarimetric mask
used to separate the two beams of polarized light produced by the
Wollaston prism in polarimetric mode. Exposure series of linear polarimetry of
the comet were taken for each filter at $\lambda$/2 retarder plate settings
of position angles (with respect to North Celestial Meridian) 0\degr,
22.5\degr, 45\degr, 67.5\degr, 90\degr, 112.5\degr, 135\degr, and 157.5\degr. 
The filter imaging and
polarimetric exposure series were obtained subsequently within a few minutes 
and less than 1 hour, respectively, per observing run. Differential 
autoguiding on a nearby star at the speed of the comet in the sky was applied during the
observations and the usual calibration exposures according to the FORS
instrument calibration plan were taken (i.e. for imaging: photometric 
standard star field, bias and sky flatfield exposures; for
the polarimetry: bias and screen flatfield exposures plus a polarized 
and unpolarized
standard star). The observing program was executed in 9 runs during 1 Oct. and
14 Dec. 2006. Details of the observations are summarized in Table \ref{obs_log}.
\begin{table*}
\begin{center}
\caption{\label{obs_log}Observing log and measurement results of the 
photometry and linear polarimetry of comet 2P/Encke. }
\begin{small}
\begin{tabular}{lrrrcrrrrrc}
\hline
\multicolumn{1}{c}{EPOCH}               &  
\multicolumn{1}{c}{PHASE}               &  
\multicolumn{1}{c}{$r$}                 &  
\multicolumn{1}{c}{$\Delta$}            &  
\multicolumn{1}{c}{Filter}              &  
\multicolumn{1}{c}{$H_0$}               &  
\multicolumn{1}{c}{\pq}                 &  
\multicolumn{1}{c}{\pu}                 &  
\multicolumn{1}{c}{\pl}                 &  
\multicolumn{1}{c}{$\zeta$}            &  
\multicolumn{1}{c}{Notes}              \\  

\multicolumn{1}{c}{(UT)}                &  
\multicolumn{1}{c}{($^\circ$)}          &  
\multicolumn{1}{c}{(AU)}                &  
\multicolumn{1}{c}{(AU)}                &  
                                        &  
\multicolumn{1}{c}{(mag)}               &  
\multicolumn{1}{c}{(\%)}                &  
\multicolumn{1}{c}{(\%)}                &  
\multicolumn{1}{c}{(\%)}                &  
\multicolumn{1}{c}{($^\circ$)}          &  
\multicolumn{1}{c}{}                   \\  
\hline\hline				 
                    &        &      &      &        &                  &                   &                  &       &              &     \\
2006-10-01 01:40:02 &   3.75 & 2.75 & 1.76 & $V$    &                  &  $-0.90 \pm 0.51$ & $ 0.03 \pm 0.51$ &  0.90 & $ 89 \pm 17$ &     \\[2mm]

2006-10-02 01:48:31 &   4.01 & 2.74 & 1.75 & $V$    &                  &  $-0.70 \pm 0.58$ & $-1.20 \pm 0.59$ &  1.39 & $120 \pm 17$ &     \\
2006-10-02 02:04:20 &   4.02 & 2.74 & 1.75 & \bcont &                  &  $-0.78 \pm 0.42$ & $-0.42 \pm 0.43$ &  0.89 & $104 \pm 19$ &     \\
2006-10-02 02:20:12 &   4.02 & 2.74 & 1.75 & $R$    &                  &  $-0.68 \pm 0.32$ & $-0.07 \pm 0.32$ &  0.68 & $ 93 \pm 15$ &     \\
2006-10-02 02:40:03 &   4.02 & 2.74 & 1.75 & \rcont &                  &  $-1.45 \pm 0.40$ & $-1.24 \pm 0.40$ &  1.91 & $110 \pm  9\phantom{1}$ & \\
2006-10-02 02:56:20 &   4.03 & 2.74 & 1.75 & $R$    & $15.16 \pm 0.07$ &                   &                  &       &              &     \\
2006-10-02 02:57:55 &   4.03 & 2.74 & 1.75 & $V$    & $15.66 \pm 0.07$ &                   &                  &       &              &     \\[2mm]
				  
2006-10-12 03:47:04 &   8.09 & 2.66 & 1.71 & $V$    &                  &  $-0.28 \pm 0.36$ & $-0.20 \pm 0.35$ &  0.35 & $107 \pm 41$ &     \\
2006-10-12 04:03:19 &   8.10 & 2.66 & 1.71 & \bcont &                  &  $-0.50 \pm 0.30$ & $ 0.13 \pm 0.45$ &  0.52 & $ 83 \pm 29$ &     \\
2006-10-12 04:19:47 &   8.10 & 2.66 & 1.71 & $R$    &                  &  $-1.12 \pm 0.32$ & $-0.52 \pm 0.33$ &  1.23 & $102 \pm 10$ &     \\
2006-10-12 04:40:12 &   8.11 & 2.66 & 1.71 & \rcont &                  &  $-0.95 \pm 0.49$ & $ 0.02 \pm 0.98$ &  0.95 & $ 89 \pm 30$ &     \\
2006-10-12 04:56:35 &   8.12 & 2.66 & 1.71 & $R$    & $15.45 \pm 0.07$ &                   &                  &       &              &     \\
2006-10-12 04:58:06 &	8.12 & 2.66 & 1.71 & $V$    & $15.93 \pm 0.09$ &                   &                  &       &              &     \\
2006-10-12 05:02:34 &   8.12 & 2.66 & 1.71 & $R$    & $15.45 \pm 0.07$ &                   &                  &       &              & [1] \\
2006-10-12 05:04:11 &   8.12 & 2.66 & 1.71 & $V$    & $15.98 \pm 0.08$ &                   &                  &       &              & [1] \\[2mm]
		      	   
2006-10-20 00:21:07 &  11.79 & 2.60 & 1.71 & $R$    & $15.68 \pm 0.07$ &                   &                  &       &              &     \\
2006-10-20 00:22:28 &  11.79 & 2.60 & 1.71 & $V$    & $16.12 \pm 0.07$ &                   &                  &       &              & [2] \\
2006-10-20 00:33:30 &  11.79 & 2.60 & 1.71 & $V$    &                  &  $ 0.07 \pm 0.42$ & $-0.36 \pm 0.42$ &  0.37 & $141 \pm 38$ &     \\
2006-10-20 00:49:18 &  11.80 & 2.60 & 1.71 & \bcont &                  &  $-0.46 \pm 0.71$ & $ 0.43 \pm 0.66$ &  0.63 & $ 68 \pm 44$ &     \\
2006-10-20 01:05:16 &  11.80 & 2.60 & 1.71 & $R$    &                  &  $-0.24 \pm 0.32$ & $ 0.05 \pm 0.31$ &  0.24 & $ 85 \pm 43$ &     \\
2006-10-20 01:25:05 &  11.81 & 2.60 & 1.71 & \rcont &                  &  $-1.54 \pm 0.82$ & $-0.01 \pm 0.67$ &  1.54 & $ 90 \pm 12$ &     \\[2mm]
 
2006-10-28 02:43:33 &  15.46 & 2.53 & 1.75 & $R$    &                  &  $-0.31 \pm 0.15$ & $ 0.08 \pm 0.15$ &  0.32 & $ 83 \pm 17$ &     \\
2006-10-28 03:18:40 &  15.47 & 2.53 & 1.75 & $V$    &                  &  $ 0.37 \pm 0.20$ & $-0.30 \pm 0.19$ &  0.48 & $160 \pm 17$ &     \\
2006-10-28 03:37:07 &  15.48 & 2.53 & 1.75 & $R$    & $15.85 \pm 0.07$ &                   &                  &       &              &     \\
2006-10-28 03:38:38 &  15.48 & 2.53 & 1.75 & $V$    & $16.17 \pm 0.07$ &                   &                  &       &              &     \\[2mm]
		      	   
2006-11-04 00:41:39 &  18.32 & 2.47 & 1.73 & $R$    &                  &  $ 0.92 \pm 0.33$ & $-0.13 \pm 0.34$ &  0.93 & $176 \pm 12$ &     \\
2006-11-04 00:59:12 &  18.32 & 2.47 & 1.73 & $V$    &                  &  $ 0.89 \pm 0.43$ & $-0.30 \pm 0.43$ &  0.94 & $171 \pm 17$ &     \\
2006-11-04 01:28:12 &  18.33 & 2.47 & 1.73 & $R$    & $15.97 \pm 0.14$ &                   &                  &       &              &     \\
2006-11-04 01:29:45 &  18.33 & 2.47 & 1.73 & $V$    & $16.40 \pm 0.22$ &                   &                  &       &              &     \\[2mm]
		      	   
2006-11-07 00:39:12 &  19.47 & 2.44 & 1.74 & $R$    &                  &  $ 0.56 \pm 0.21$ & $-0.14 \pm 0.19$ &  0.58 & $173 \pm 12$ &     \\
2006-11-07 01:14:18 &  19.47 & 2.44 & 1.74 & $V$    &                  &  $ 1.13 \pm 0.25$ & $-0.12 \pm 0.25$ &  1.13 & $177 \pm  7\phantom{1}$ & \\
2006-11-07 01:34:45 &  19.47 & 2.44 & 1.74 & $R$    & $15.97 \pm 0.10$ &                   &                  &       &              &     \\
2006-11-07 01:36:19 &  19.47 & 2.44 & 1.74 & $V$    & $16.58 \pm 0.10$ &                   &                  &       &              &     \\[2mm]
		      	   
2006-11-10 00:56:40 &  20.54 & 2.42 & 1.75 & $R$    &                  &  $ 1.14 \pm 0.22$ & $-0.02 \pm 0.22$ &  1.14 & $179 \pm  6\phantom{1}$ & \\
2006-11-10 01:32:37 &  20.55 & 2.42 & 1.75 & $V$    &                  &  $ 0.55 \pm 0.25$ & $ 0.32 \pm 0.23$ &  0.64 & $ 15 \pm 15$ &     \\
2006-11-10 01:53:11 &  20.55 & 2.42 & 1.75 & $R$    & $15.98 \pm 0.08$ &                   &                  &       &              &     \\
2006-12-14 01:08:21 &  27.97 & 2.09 & 1.94 & $R$    &                  &  $ 1.90 \pm 0.40$ & $-0.37 \pm 0.41$ &  1.93 & $175 \pm  7\phantom{1}$ & [3] \\
2006-12-14 01:15:21 &  27.97 & 2.09 & 1.94 & $V$    &                  &  $ 1.20 \pm 0.96$ & $-1.51 \pm 1.12$ &  1.93 & $154 \pm 22$ & [3] \\
\hline
\end{tabular}
\end{small}
\end{center}
{\it Explanations:} The table lists the observing epoch (midpoint of exposure series), 
the Sun (r) and Earth ($\Delta$)
distances, the phase angle ($\phi$), the filter used, the absolute filter
brightness m(1,1,$\phi$) for r = $\Delta$ = 1 AU and phase angle $\phi$, 
the Stokes parameters \pq\ and \pu, the fraction of linear polarization
\pl\ and the angle $\zeta$ of maximum polarization.
Comments are indexed and explained below:
[1] Photometric exposures in $V$ and $R$ were repeated because of guide probe vignetting.
[2] No zeropoints available in $V$ on 2006-10-20. We used those obtained on 2006-10-19.
[3] Target very close to a bright star, accurate polarimetry and photometry
impossible.
\end{table*}

\subsection{Reduction of the Photometry Data}\label{photreduction} 
All images were bias subtracted, and images used for photometry were 
then divided by a master flat field obtained from four sky flats 
taken at twilight.  For the background subtraction, we subtracted 
in a first step a constant value that was 
estimated in regions far from the comet photometric center, where the 
contribution of a possible coma was negligible. A 
further correction was evaluated and subtracted by measuring the $\Sigma$Af 
function, defined by Tozzi et al. (2004), versus the projected nucleocentric 
distance, $\rho$. The $\Sigma$Af function is defined by the product of 
the geometric albedo multiplied 
with the total area covered by the solid component (usually the cometary dust;
A'Hearn et al. 1995)
in an annulus of radius $\rho$ and unitary depth (Tozzi et al. 2004). 
For a normal comet, the $\Sigma$Af function is constant,
 and in the case of no coma it should be zero. Following a trial and error 
procedure, different background values were subtracted until the $\Sigma$Af 
function was measured to be 
constant versus $\rho$. As 
shown in Figure~2, $\Sigma$Af shows a peak at the cometary center, where the 
light-scattering contribution from the nucleus dominates, and decreases to values close to zero at larger projected distances from the nucleus 
(for a more detailed discussion see below). Finally, 
the images were calibrated both in magnitude and in 
Af. The value Af describes the product of the mean albedo A and 
the filling factor f of the dust grains in a measurement aperture of 
radius $\rho$. The product Af$\rho$ is a measure of the dust production of the
comet (for details on Af$\rho$ see A'Hearn et al. 1995). 
Using standard star images we determed photometric zeropoints
for the respective observing nights assuming atmospheric extinction
coefficients of 0.114 and 0.065 mag/airmass and instrument color coefficients 
of 0.03 and 0.06 (plus solar colors for the comet to a first approximation)
for $V$ and $R$ filters, respectively. The final two
photometric parameters could not be determined from the available calibration
images directly and were adopted after a careful analysis of the respective
information on ESO's data quality information web page for the FORS
instrument\footnote{ see http://www.eso.org/observing/dfo/quality}. 
The cometary photometry was measured using a constant
aperture diameter of 6''.

\subsection{Reduction of the Linear Polarimetry Data}\label{polreduction} 
All polarimetric images were bias subtracted and flatfield corrected. 
For the background subtraction, we followed 
the same strategy used for the imaging data of the comet, by 
treating the ordinary and extra-ordinary beams separately. 
A further refinement in the 
background estimate was obtained by measuring (for each position 
angle of the retarder waveplate) the ratio
\begin{equation}
h = \frac{(\fo - \fe)}{(\fo + \fe)},
\end{equation}
where \fo\ is the flux in the ordinary beam and \fe\ the flux in the 
extraordinary beam, in annuli of increasing radius. We then 
readjusted the value used for background subtraction until the 
measured $h$ values became independent of the projected 
nucleocentric distance $\rho$. In the $R$ band, typically, 
the background values obtained 
after the first step (i.e. estimated from regions far from the 
cometary nucleus) were approximately 800 e$^-$; the first correction,
based on the analysis of the $\Sigma$Af function, was approximately 10 
- 20 e$^-$; the final correction derived by insisting that the 
ratio $h$ was constant with $\rho$, was approximately $1-2$ e$^-$.

Linear polarization was then calculated from
\begin{equation}
q' = \frac{2}{N} \sum_{i=0}^N r(\alpha_i) \cos(4\alpha_i),
\end{equation}
and
\begin{equation}
u' = \frac{2}{N} \sum_{i=0}^N r(\alpha_i) \sin(4\alpha_i),
\end{equation}
where $q'$ and $u'$ are the reduced Stokes parameters defined 
according to Shurcliff (1962) and measured using as the reference direction 
the North Celestial Meridian, and $\alpha_i = 22.5\degr \times i$ are 
the position angles of the retarder waveplate. From $q'$ and $u'$, we 
finally calculated the linear polarization $\pq\ = Q/I$ and $\pu = U/I$, 
by adopting as the
reference direction the perpendicular to the great circle 
passing through the comet and the Sun at the observing epoch, 
as explained in Landi Degl'Innocenti et al. (2007). 
\pq\ and \pu\ allow one to determine
the total fraction of linear polarization $\pl = \sqrt{\pq^2 + \pu^2}$ and the angle of
maximum polarization $\zeta$.
\pq\ represents the flux perpendicular to the 
plane Sun-Comet-Earth (the scattering plane) minus the flux parallel 
to that plane, divided by the sum of the two fluxes, and $\zeta$ is 
the angle between the perpendicular to the scattering 
plane and the direction of maximum polarization. 
As for the photometry, a fixed aperture of 15'' diameter was
applied for the polarimetric measurements of the comet. In the presence of
coma, we have implicitly assumed that the polarization 
of the coma does not depend on $\rho$ (see below for a discussion of the 
influence of a faint coma on the polarization data of the comet). 


\section{Results}\label{results}
\begin{figure}[ht]
\centering
\vspace*{0.3cm}
\resizebox{!}{5cm}{\includegraphics{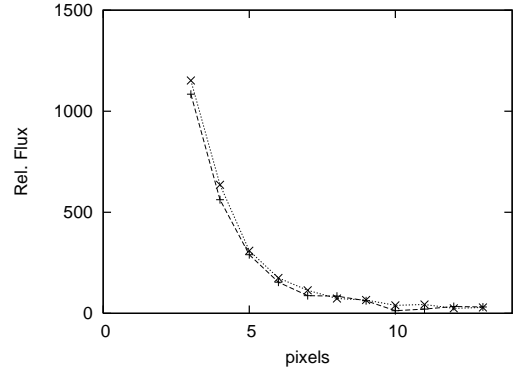}}
\caption{\label{fig_PSF}
Azimuthally averaged PSF profile of the comet (dashed line) compared with that
of a background star in the same exposure. The image was taken on 28 Oct. 2006
through the $V$ filter.}
\end{figure}

{\bf Search for coma:} No coma is detectable by direct visual inspection of the comet 
images. To confirm the presence of a faint coma possibly
undetectable by visual inspection and measure its contribution to the
object brightness, two numerical analysis methods were applied.
The first and classical method (see for instance Boehnhardt et al. 2002) compared the azimuthally 
averaged profiles of the comet with the instrumental point spread function
(PSF) measured from background stars in the same exposure. Since the comet 
had a proper motion and the telescope was tracking the comet, 
the stars appeared as short trails. Hence, the stellar PSF can only be measured 
in one direction i.e. perpendicular to the cometary motion and
averaged along the trail direction. Figure \ref{fig_PSF} shows that the stellar and
cometary PSF are almost identical, in particular in the PSF wings and close to the
background level at which the coma light is expected to provide the most
significant contribution.

To increase the sensitivity to coma detection, all polarimetric images recorded
in a single filter during a single observing night were centered on the
comet position and coadded, reaching a total integration time of
approximately 1000s
(instead of the short 60s exposures used for normal photometric observations). 
Due to seeing variations during exposure series, direct comparison of the
PSFs of comet and background stars was complicated and usually impossible. 
For these composite images a second analysis method was therefore
applied: Measurement of the $\Sigma$Af as function of $\rho$ (see above).
In the case of the absence of coma, this function should approach zero at radial
distance $\rho$ of about 3 FWHM from the central brightness peak, 
with FWHM being the value of the full width at half maximum of the 
PSF of the coadded comet image. In the case of the coadded polarimetric images
of 2P/Encke, all $\Sigma$Af profiles show small, but significant non-zero
fluxes at larger distances from the center of the coadded comet
image. We recall that the background level has been subtracted
using $\Sigma$Af as described above and that any residual background 
level would result in a (likely linear) non-constant trend
of $\Sigma$Af versus $\rho$. The fact that for the 2P/Encke polarimetric images 
$\Sigma$Af is constant with $\rho$ means that the intensity profile of the
comet declines with 1/$\rho$, a phenomenon that is typical of an expanding dust
coma around the nucleus. 
We therefore conclude that a weak coma was present around the comet during
the entire period of our observations, i.e. from early Oct. to mid Dec. 2006
when the comet was approaching the Sun from 2.75 to 2.1 AU. Given the
overall $\Sigma$Af profile, it is unlikely that the central brightness
peak of the comet images is due to an unresolved dense dust coma; instead, we
consider the photometric flux of the peak produced by surface-reflected
sunlight.

\begin{figure}[h]
\centering
\vspace*{0.3cm}
\resizebox{!}{5cm}{\includegraphics{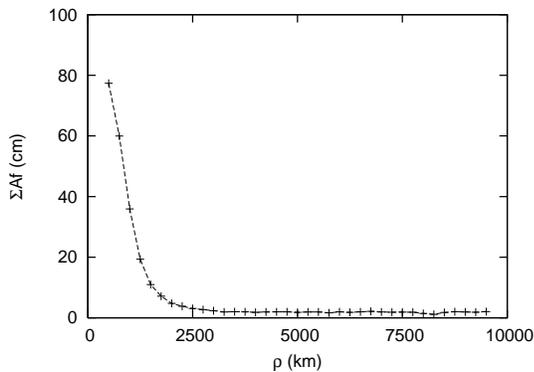}}
\caption{\label{fig_SAf}
$\Sigma$Af profile as function of the nucleocentric distance $\rho$,
obtained from polarimetric observations in $V$ filter on 28 Oct. 2008. The total 
exposure time is 960s. For radial distances greater 
than 2000 km, $\Sigma$Af is small but different from the background level of
zero flux. Hence a very faint coma is present around the cometary nucleus.}
\end{figure}

Flux calibration of the polarimetric images was obtained by comparing the
inner part of the
$\Sigma$Af profile of the coadded polarimetric image with that of a calibrated 
normal filter exposure of the comet taken in the same filter during the same night. 
Figure \ref{fig_SAf} shows an example of the $\Sigma$Af profile of a coadded
polarimetric image series of comet 2P/Encke. Within  the error margins, 
the $\Sigma$Af profiles obtained for the various observing nights show no trend,  
neither with heliocentric distance nor with phase angle. The average level 
of the weak coma flux corresponds to Af$\rho$ values (for the definition 
of Af$\rho$ see above) of 0.65$\pm$0.35 cm in $V$ and 
0.49$\pm$0.19 cm in $R$ . They are equivalent to a dust
production rate Q$_{dust}$ of the order of about 0.05 kg/s, assuming a simple
empirical relationship between Af$\rho$ and Q$_{dust}$ used 
by Kidger (2004). The measured Af$\rho$ values equal no more than 1 percent 
of the total cometary signal and its effect on the photometric and
polarimetric data analysis of 2P/Encke described below, can be neglected. 
We also note that during the
reduction of the polarization images a weak extended coma was subtracted as part of
the general background signal measured beyond 3 FWHM of the PSF of the
comet. Only second order contributions to the $Q$ and $U$ measurements of
the comet signal may remain from polarized light of the dust coma around the
nucleus. Since estimated to be below 1 percent of the measured Stokes
parameters, we ignored these in the overall error analysis of the
polarimetric results of 2P/Encke.

\begin{figure}[h]
\includegraphics[height=88mm,width=70mm,angle=-90,clip=true] {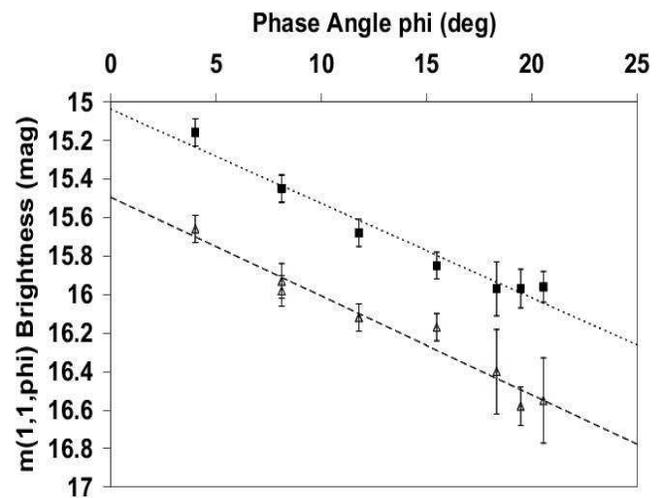}
\caption{\label{phot_phasefunction} The brightness phase function of the
  nucleus of comet 2P/Encke in broadband $V$ and $R$ filters. The plot
  shows the measured m(1,1,$\phi$) brightness, in addition to the linear fit 
  functions wersus phase angle. Filled squares = $R$ filter, open 
  triangles = $V$ filter.}
\end{figure}

{\bf Photometric phase function of the nucleus:} Figure \ref{phot_phasefunction}
shows the photometric phase function of the nucleus of comet 2P/Encke for
broadband $V$ and $R$ filters. The values plotted therein, the so-called 
m(1,1,$\phi$) magnitudes (see Table \ref{obs_log}), are derived 
from the observed filter magnitudes by removing the dependencies 
on the Sun and Earth distances for the respective observing epochs. The
figure shows a steep almost linear brightness increase for the nucleus 
with decreasing phase angle. It also suggests that there is - at least 
in the $R$ filter data - a slight non-linear increase in brightness
toward zero phase, starting all the way from $\phi$ = 20\degr. Although this
deviation from linear slope is within the expected amplitude of
brightness variations due to the rotation of the non-spherical nucleus, its
systematics suggest that the opposition effect may play a role in the nucleus 
phase function of 2P/Encke. Ignoring the possible minor non-linearity in the
phase functions of the comet, we found almost identical slopes $\beta$ when
fitting a linear phase curve, i.e. for the $V$ filter data 
$\beta$ = 0.051$\pm$0.004 mag/\degr
and for the $R$ filter ones $\beta$ = 0.049$\pm$0.004 mag/\degr. Our $\beta$ is 
slightly smaller, although consistent within the error margins, 
than the linear slope parameter ($\beta$ = 0.06 mag/\degr) 
determined by Fernandez et al. (2000) from a compilation of
published nucleus magnitudes of the comet measured over a wider phase-angle
range and that found by Jewitt (2004) for 2P/Encke ($\beta$ = 0.060$\pm$0.005 
mag/\degr) derived from a much sparser dataset.
We conclude that the phase function of 2P/Encke follows a
color-neutral almost linear brightening law with $\beta$ = 0.050$\pm$0.004 mag/\degr, 
which is slightly steeper than the canonical value ($\beta$ = 0.04mag/\degr; Lamy et
al. 2004), although within the range found for other cometary nuclei (Jewitt
2004, Lamy et al. 2004).  
   
{\bf Nucleus size and colors:} Extrapolation of the linear phase function of
2P/Encke to zero phase angle provides the absolute magnitude m(1,1,0) of the
nucleus, i.e. 15.50$\pm$0.06 and 15.03$\pm$0.05 mag for $V$ and $R$ filter,
respectively. With the geometric albedo of 0.047 (Lamy et al. 2004;
Fernandez et al. 2000), we obtain an equivalent radius of 
2.43$\pm$0.06km in $R$, in good agreement with the radius estimate 
by Fernandez et al. (2000; 2.4$\pm$0.3km), Kelley et al. (2006), and Campins et
al. (1988) and slightly
smaller than radar echo results by Harmon and Nolan (2005; 2.42 - 3.72km). 
We emphasize that our results may be affected by insufficient sampling
of the rotation light curve and may deviate from earlier findings
because of different viewing aspects of the nucleus along the
orbit and with time (see Belton et al. 2005). The {\it V-R} color of the nucleus is
0.47$\pm$0.07 mag, which corresponds to an intrinsic color (i.e. corrected for
solar $V$ - $R$ color) of 0.11$\pm$0.07 mag or 
a spectral slope S' of 11$\pm$8 \%/100nm. Our mean {\it V-R} color and the 
spectral gradient S' are slightly
higher than that measured by Jewitt (2004) and Lowry \& Weissman (2007),
although still within the error margins. 

\begin{figure}[h]
\includegraphics[angle=-90,width=88mm,clip=true] {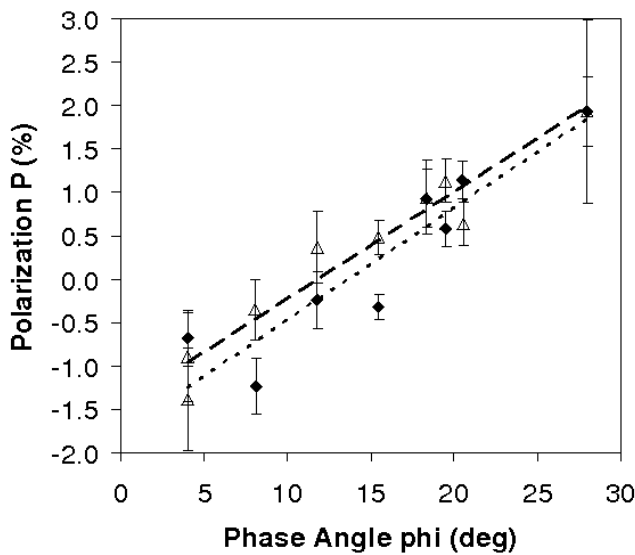}
\caption{\label{pol_phasefunction} Linear polarization P of comet 2P/Encke
  versus phase angle. Filled triangles/squares = V/R filter measurements,
  long/short dash lines = linear fits to V/R filter measurements.}
\end{figure}

{\bf Polarimetric phase function of the nucleus:} Stokes $P_Q$ and $P_U$ of the linear
polarimetry measured for 2P/Encke are compiled in Table \ref{obs_log}. The $P_Q$
and $P_U$ values listed refer to the scattering plane Sun-comet-observer. 
The polarization $P_L$ with respect 
to the scattering plane (listed in
Table \ref{obs_log}) is plotted versus phase angle in
Figure \ref{pol_phasefunction} for the two broadband filters used (V
and R). In both broadband filters the polarization shows the 
same, linear 
increase with increasing phase angle $\phi$ 
(from about -1\% at $\phi$=4\degr to about 
+1.7\% at $\phi$=28\degr). The minimum polarization values for $V$ and $R$ 
correspond to the upper limits only,
since our polarization phase functions do not show the expected turn-over
towards zero polarization for small phase angles (the minimum polarization is
rather the lower end of the linear phase function slope and the turn-over
occurs at $\phi<4$\degr only). Zero polarization 
is passed at about 13\degr  phase angle (inversion angle). The slope of 
the polarization increase is determined by linear regression 
to be $0.123\pm0.013$ \%/\degr in $V$ and $0.129\pm0.021$ \%/\degr 
in {\it R}, i.e. within the errors the phase-angle gradients are identical and 
do not change with filter.

From the values listed in Table \ref{obs_log}, it is obvious that for small $P_Q$,
i.e. close to the overall measurement uncertainty, Stokes $P_U$
has comparable amplitudes to $P_Q$ which produces a considerable scatter and large
uncertainties in the values for the position angle $\zeta$ 
of maximum linear polarization. 
The position angle $\zeta$ is well determined for higher polarization 
values and agrees well with the position angle of the light scattering plane. 
However, the close-to-zero level of
Stokes $P_U$ for all measurements provides evidence for the correctness of the
data reduction, and is in agreement with theoretical expectations for a
light-scattering-plane oriented polarization (see Muinonen et al. 2002). 
For the narrowband measurements using filters \bcont\ and \rcont, Stokes $P_Q$ is
systematically higher for the red continuum filter compared with the blue,
although all results are affected by relatively large errors. Within the
margins of the
errors the measured Stokes parameters agree with the Stokes parameters
measured for the $V$ and R filters, respectively. 

To our knowledge, 2P/Encke is the only comet that has linear polarization
at the nucleus measured. Jewitt (2004) published a set of
polarization values (1.5, 1.0, 1.9\% $\pm$0.5 \%), obtained on 22 Oct. 
2003 at phase angle 22.1\degr  through cometary dust continuum filters 
(central wavelengths of 445.3, 525.9, 713.3 nm). His results are consistent 
with our own.

\begin{table*}
\caption{\label{comparison} Comparison of photometric and polarimetric 
properties of comet Encke and other small bodies.}
\begin{tabular}{lcccccccl}
\hline
Object & Albedo  & Phase coefficient,                 &  Spectral    &
$|P_{min}|$ & $\phi_{min}$ & Pol. slope & Pol.Inv. & Pol.spect.\\
       &  $p_V$   & $ \beta$ for  $5 \le \phi \le 30^o$& Slope &
(\%)       & (\degr)       & (\%/\degr)  & angle $\phi_o$      & gradient\\
       &          & (mag/\degr)                         & (\%/100nm)    &
           &              &            & (\degr)        & at  $\phi \le 30^o$\\
\hline
2P/Encke&0.047&0.048 &11&$\ge$ 1.4&$\le$4&0.123&13& negative\\
        &     &      &    &         &      &     &  & decreases with $\phi$\\
Comet nuclei&0.02--0.06 &0.043--0.059  & 0--41 &?&?&?&?& ?\\
        &     &      &    &         &      &     &  & \\
Comet dust&0.05&0.018--0.035 &-1--12&1.5&9.6&0.31&21.8&negative\\
        &     &      &    &         &      &     &  & decreases with $\phi$\\
2060/Chiron &0.07 &0.045 & 1 &~1.5&1.5--2&Out of range&~9 (?) & negative ?\\
        &     &      &    &         &      &     &  & \\
C asteroids&0.02--0.09&0.042&  -3--20&1.73&9.4&0.28&20.5&positive\\
        &     &      &    &         &      &     &  & increases with $\phi$\\
S asteroids&0.085--0.33&0.029&  -15--88&0.77&6.3&0.09&20.1&negative\\
        &     &      &    &         &      &     &  & increases with $\phi$\\
E asteroids&0.34--0.6&0.019&rather flat spectra&0.3&5.8&0.04&17.8&\\
        &     &      &    &         &      &     &  & \\
M asteroids&0.074--0.25&0.033& 6--15&0.96&7.7&0.09&23.5&negative\\
        &     &      &    &         &      &     &  & increases with $\phi$\\
F asteroids&0.024--0.086&0.044& -24 
(Interamnia)&1--1.4&7.6--8&0.35--0.4&14--17&slightly negative or\\
        &     &      & bluish spectra   &   &      &     &  & neutral\\
\hline
\vspace{.3mm}
\end{tabular}
{\it Explanations:} Column ``Spectral Slope'' 
lists the slope of the visible spectrum of the objects as determined 
from measured filter colors corrected for the color of the Sun 
(see Boehnhardt et al. 2001); ``$|P_{min}|$'' gives
the value of minimum polarization, ``$\phi_{min}$'' the phase angle of
$|P_{min}|$, ``Pol.slope'' the gradient of the polarimetric phase function and
``Pol.Inv.'' the polarization inversion angle, i.e. when the sign of
polarization $P_L$ changes from negative to positive. Column
``Pol.spect. gradient at $\phi \le 30^o$'' describes the overall spectral
shape of the object spectra for phase angles below 30 \degr. 
The references for the table values are: albedo for comet 
nuclei - Lamy et al 2004; filter colors of objects - Snodgrass et al. 2008;  
phase coefficients from space missions - Soderblom et al. 2002, Li et al. 2006
phase coefficients from ground-based observations - Jewitt and Sheppard
2004, Jewitt et al. 2003; albedo for 2060/Chiron - Groussin et al. 2004; $V$ -
$R$ color of 2060/Chiron - Davies et al. 1998; the rest of 2060/Chiron's  data -
 Bagnulo et al. 2006; asteroid albedos - Tedesco et al. 1989; asteroid phase 
coefficients - Bowell et al. 1979; asteroid filter colors - Hansen 1976, Chapman
\& Gaffey 1979; polarimetric asteroid data - Lupishko 2006; spectral gradients 
of polarization for asteroids - Belskaya et al. 2008; comet dust data - 
Kolokolova et al. 2004, Kiselev \& Rosenbush 2004, Meech \& Jewitt
1987. Question marks indicate that no comparison data are available.
\end{table*}

{\bf Test of the asteroid polarization-albedo relationship:} We also
applied the polarization-albedo relationship for asteroids 
(Zellner and Gradie 1976, Zellner et al. 1977) to the 2P/Encke polarimetry
data to validate this empirical method of albedo estimation for the nucleus 
of the comet. Using the phase
function parameters of Lupishko and Mohamed (1996) we derive a geometric 
albedo of 0.145 (range 0.13--0.16), when using the average slope of the 
polarization phase function (see above), and of 0.08 (range 0.06-0.16) 
when using the values of minimum polarization P$_{min}$ - which by 
itself is an upper limit only. Unfortunately, in our measurements of 2P/Encke
P$_{min}$ is not well determined and the upper limit used instead shows  
relatively large errors (see
Table \ref{obs_log}). In conclusion, the empirical polarization-albedo
relationships for asteroids suggests a significantly higher geometric 
albedo for the nucleus of 2P/Encke compared to that obtained from the more
realistic approach of visible and thermal IR measurements of the nucleus 
(0.047; Lamy et al. 2004; Fernandez et
al. 2000). This may indicate that either the empirical polarization-albedo
relationship of asteroids does not apply to cometary nuclei, or at least not
for 2P/Encke, or, if applicable, that it requires different 
fitting parameters than determined from asteroids. It is indeed been proposed
that the empirircal polarization-albedo rule is not applicable to extremely
dark objects such as cometary nuclei (Dollfus \& Zellner 1979). 
The negative outcome of
this test may imply that the surface constitution of 2P/Encke 
differs from those of main-belt and near-Earth asteroids for which the 
empirical relationship is calibrated by measurements.


\section{Discussion and Conclusions}
\label{discussion}

We have presented photometric and linear polarimetric phase functions of 2P/Encke 
measured during the 2006 approach of the comet to the Sun. The weak coma 
found about the comet does not affect significantly the photometric and 
polarimetric light-scattering phase functions measured, i.e. the 
phase functions reflect the light-scattering behavior
of the nucleus surface. Both phase functions show an almost linear behavior with phase
angle $\phi$ (over the measured range from 4 to 28 deg). Trends in the phase function
with wavelength are not obvious for the nucleus photometry, but may exist for
the linear polarization. A small opposition surge is tentatively 
suggested in the photometric data, but needs verification by further
observations at smaller phase angles. Similarly, the value of minimum polarization 
P$_{min}$ in the polarization phase curve should be reassessed by new
observations covering the range to zero phase angle that is of immediate
interest for modeling the opposition effects in the light-scattering by the
cometary nucleus.   
Our radius estimation and spectral slope S' agree with earlier measurements, 
in particularly well with those obtained from 
IR observations. The
low activity level of the nucleus at solar distances of 2.75 to 2.1AU suggests
that the cometary nucleus is covered by a crust that prevents major
outgassing. This appears to be true during this part of the orbit, even though
2P/Encke has entered the distance range where water sublimation usually
generates significant cometary activity. 

Table \ref{comparison} compares light-scattering properties of the 
nucleus of comet 2P/Encke with those of other solar-system
bodies, that may have similar surfaces and were observed in similar
conditions (phase angle, wavelength).  The majority of the data are for V
filter - if not indicated otherwise in the table - at the phase angles
5-30\degr. The analysis of Table \ref{comparison} indicates that comet 
2P/Encke displays photometric properties
typical of cometary nuclei. As for other cometary nuclei, 
the slope of the phase function differs from that measured for cometary 
dust, which might be due to the difference between light scattering on individual dust particles
and their compact layers at the nuclear surfaces. 

Among asteroids, the light scattering behavior of C-type
objects resembles most closely that of 
comet 2P/Encke, at least its photometric properties. There are major
differences between the polarization phase curve 
(Figure \ref{pol_phasefunction}) of 
the nucleus of 2P/Encke and the polarization phase curves of asteroids of 
different taxonomy. Primitive asteroids of type C, P, G, and B and 
evolved asteroids of type V, S, M, E, and A have, in general, larger 
inversion angles than 2P/Encke (cf. Muinonen et al, 2002; 
Fornasier et al., 2006), while only primitive F-type asteroids 
exhibit similar inversion angles (Belskaya et al., 2005). 
However, other polarimetric characteristics of F-type
asteroids differ significantly: The angle of the polarization minimum is 
at least twice as
large, and the slope of polarization at the inversion angle is almost three
times larger than that of comet 2P/Encke. Moreover, F-type asteroids do not 
exhibit red colors, but display neutral or even bluish surface 
colors. 
Unfortunately, we cannot compare the properties of comet 2P/Encke 
with icy bodies such as Transneptunian objects or the small 
satellites of Saturn, Uranus, and Neptune since these
objects can be observed only at small phase angles. The Centaur
2060/Chiron may be the object most appropriate for the comparison. 
Compared to comet 2P/Encke, it has however neither similar photometric
(too high albedo, too red color) nor polarimetric data (too small inversion  
angle). 
The polarimetry therefore reveals unique properties of the nucleus of comet
2P/Encke that may be typical for cometary nuclei but differ significantly
from the properties of other solar-system bodies. This uniqueness does not
manifest itself photometrically. It may indicate a specific composition or
structure of the surface layer of cometary nuclei. 
Whether the narrow negative polarization branch can be explained qualitatively
by the single-scattering interference mechanism (Muinonen et al.
2007) remains to be answered by a future study. Based on that mechanism, negative
polarization branches of icy objects which corresponds to smaller real parts of refractive
indices, can be expected to be narrower than those of silicate-rich stony
objects, which represents larger real parts of refractive indices.

The polarization minimum of 2P/Encke, which is not well constrained 
by the present data set, is less than or equal to -1\% and, as mentioned above,
it is possibly within the range of data for F-type asteroids 
that have polarization minima between about -1.0 and -1.5\% 
(Belskaya et al. 2005). However, we reiterate that the polarization phase
curve of 2P/Encke is the only data set for comets available to ourselves.
The number of F-type asteroids with measured
polarization phase curves is also small such that, in terms of polarization
properties, it is difficult to decide whether comets and F-type asteroids 
are similar. 

Our 2P/Encke results have shown that the empirical albedo-polarization
relationships of asteroids cannot easily be applied to cometary nuclei,
possibly because the surface constitution and light-scattering properties
are different. Whether such a relationship can be established
for cometary nuclei remains undecided until more objects are studied and a
self-consistent modeling of photometric, spectroscopic, and polarimetric 
light-scattering properties for comets is available.

\acknowledgement{We would like to thank everyone at the European Southern
  Observatory at Cerro Paranal in Chile and in Garching/Germany,
  who were involved in the
  implementation and execution of this service mode program at the Very Large
  Telescope VLT.}

\end{document}